\documentclass[aps,prl,twocolumn,amsmath,amssymb]{revtex4-1}
\usepackage{color}
\usepackage{graphicx}
\usepackage{hyperref}
\usepackage{txfonts}
\usepackage{xcolor}

\makeatother

\begin{document}

\title{
Randomisation of Pulse Phases for Unambiguous and Robust Quantum Sensing
}

\author{Zhen-Yu Wang$^{1,\dagger}$}
\email{E-mail: zhenyu.wang@uni-ulm.de}
\author{Jacob E. Lang$^{2}$}
\thanks{These authors contributed equally to this work}
\author{Simon Schmitt$^{3}$}
\thanks{These authors contributed equally to this work}
\author{Johannes Lang$^{3}$, \\Jorge Casanova$^{4,5}$, Liam McGuinness$^{3}$, Tania S. Monteiro$^{2}$, Fedor Jelezko$^{3}$}
\author{Martin B. Plenio$^{1}$}
\affiliation{1. Institut f\"ur Theoretische Physik und IQST, Albert-Einstein-Allee 11, Universit\"at Ulm, D-89069 Ulm, Germany}
\affiliation{2. Department of Physics and Astronomy, University College London, Gower Street, London WC1E 6BT, United Kingdom}
\affiliation{3. Institute of Quantum Optics, Albert-Einstein-Allee 11, Universit\"at Ulm, D-89069 Ulm, Germany}
\affiliation{4. Department of Physical Chemistry, University of the Basque Country UPV/EHU, Apartado 644, 48080 Bilbao, Spain}
\affiliation{5. IKERBASQUE, Basque Foundation for Science, Maria Diaz de Haro 3, 48013, Bilbao, Spain}

\begin{abstract}
We develop theoretically and demonstrate experimentally a universal dynamical decoupling method for robust quantum sensing with unambiguous signal identification. Our method uses randomisation of control pulses to suppress simultaneously two types of errors in the measured spectra that would otherwise lead to false signal identification. These are spurious responses due to finite-width $\pi$ pulses, as well as signal distortion caused by $\pi$ pulse imperfections. For the cases of nanoscale nuclear spin sensing and AC magnetometry, we benchmark the performance of the protocol with a single nitrogen vacancy centre in diamond against widely used non-randomised pulse sequences. Our method is general and can be combined with existing multipulse quantum sensing sequences to enhance their performance.
\end{abstract}

\maketitle

\emph{Introduction.--}
The nitrogen-vacancy (NV) centre~\cite{doherty2013} in diamond has demonstrated excellent sensitivity and nanoscale resolution in a range of quantum sensing experiments~\cite{schirhagl2014nitrogen,rondin2014magnetometry,suter2016single,wu2016diamond}. In particular, under dynamical decoupling (DD) control~\cite{souza2012robust} the NV centre can be protected against environmental noise~\cite{ryan2010robust,deLange2010universal,naydenov2011dynamical} while at the same time being made sensitive to an AC magnetic field of a particular frequency~\cite{deLange2011single}. This makes the NV centre a highly promising probe for nanoscale nuclear magnetic resonance (NMR) and magnetic resonance imaging (MRI)~\cite{staudacher2013nuclear,rugar2015proton,deVience2015nanoscale,shi2015single,schmitt2017sub,boss2017quantum,glenn2018high,rosskopf2017quantum,pham2016nmr}. Moreover, NV centers under DD control can be used to detect, identify, and control nearby single nuclear spins~\cite{taminiau2012detection,kolkowitz2012sensing,zhao2012sensing,Muller2014,lang2015dynamical,sasaki2018determination,zopes2018nc,zopes2018prl,pfender2019high} and spin clusters~\cite{zhao2011atomic,shi2014sensing,wang2016positioning,wang2017delayed,abobeih2018one}, for applications in quantum sensing~\cite{degen2017quantum}, quantum information processing~\cite{casanova2016noise,casanova2017arbitrary}, quantum simulations~\cite{cai2013a}, and quantum networks~\cite{humphreys2018deterministic,perlin2019noise}.

Errors in the DD control pulses are unavoidable in experiments and limit performance especially for larger number of pulses. To compensate for detuning and amplitude errors 
in control pulses, robust DD sequences that include several pulse phases~\cite{gullion1990new,ryan2010robust,casanova2015robust,genov2017arbitrarily} were developed. However, these robust sequences still 
require good pulse-phase control and, more importantly, they introduce spurious harmonic response \cite{loretz2015spurious} due to the finite length of the control pulses. This spurious 
response leads to false signal identification, e.g. the misidentification of $^{13}$C nuclei for $^{1}$H 
nuclei, and hence impact negatively the reliability and reproducibility of quantum sensing experiments.
Under special circumstances it is possible to control some of these spurious peaks~\cite{haase2016pulse,lang2017enhanced,shu2017unambiguous}.  
However, it is highly desirable to design a systematic and reliable method to suppress any spurious response and to improve robustness of all existing DD sequences, such as the routinely used XY family of sequences~\cite{gullion1990new}, the universally robust (UR) sequences~\cite{genov2017arbitrarily}, and other DD sequences leading to enhanced nuclear selectivity~\cite{casanova2015robust,haase2018soft}.

\begin{figure}
\center
\includegraphics[width=1.0\columnwidth]{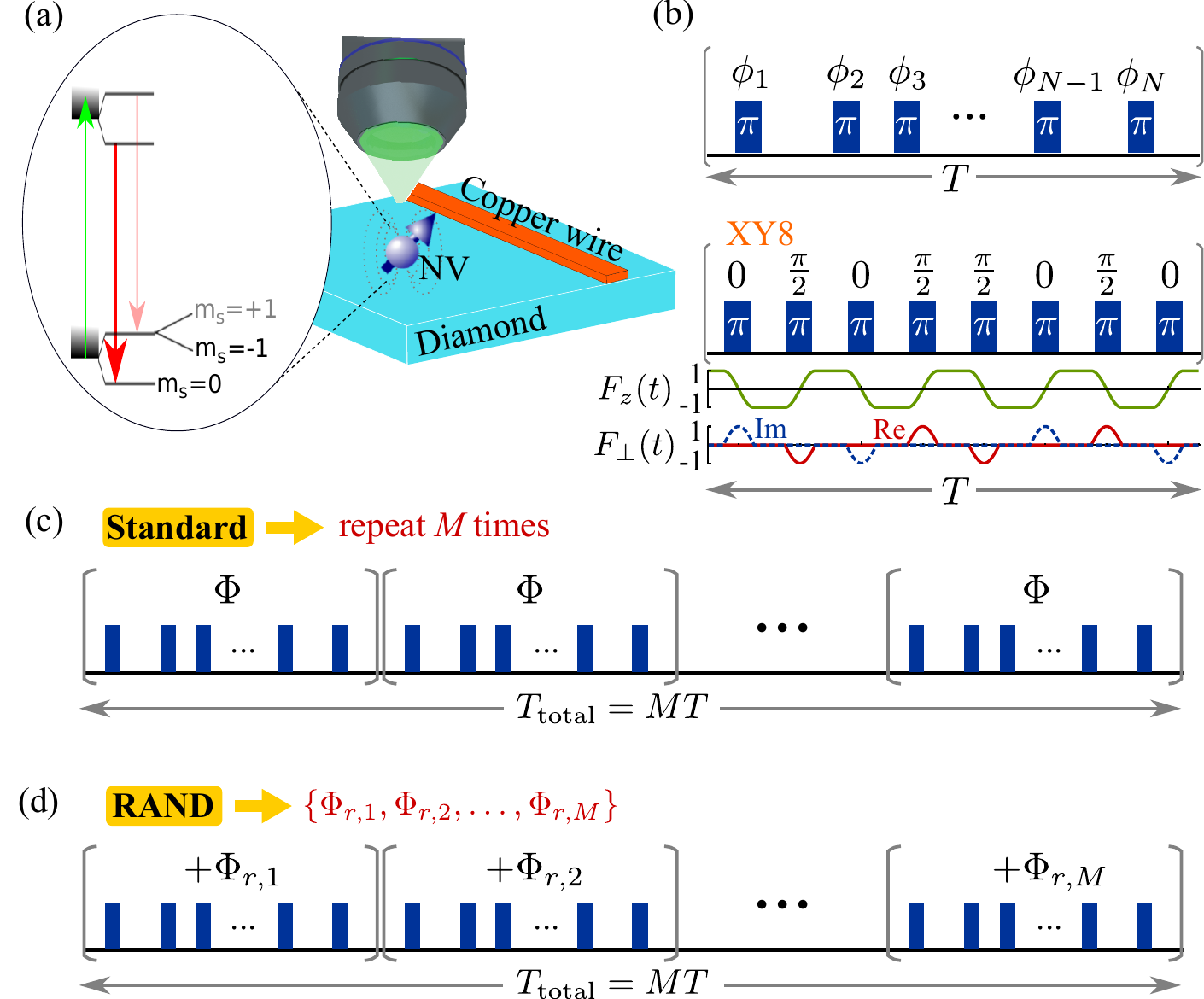}
\caption{Randomisation protocol for quantum sensing. (a) Experimental set-up with an NV centre in diamond used as a quantum sensor. (b) A basic unit of pulse sequence for quantum sensing, which is defined by the positions and phases of the $\pi$ pulses. The lower panel is the example of an XY8 sequence with its associate $F_{z}(t)$ and $F_{\perp}(t)$.  (c) The standard way to construct a longer sensing sequence is to repeat the same basic pulse unit in (b) $M$ times. (d) 
The randomisation protocol shifts all the pulses within each unit by a common random phase $\Phi_{r,m}$. The random phases $\Phi_{r,m}$ at different blocks are independent. 
One may refresh all the random phases $\{\Phi_{r,1},\Phi_{r,2},\ldots, \Phi_{r,M}\}$ at different runs of the sensing experiment.}\label{Fig1}
\end{figure}

In this Letter, we demonstrate that phase randomisation upon repetition of a basic pulse unit of DD sequences is a generic tool that improves their robustness and eliminates spurious response whilst maintaining the desired signal. This is achieved by, firstly, adding a global phase to the applied $\pi$ pulses within one elemental unit and, secondly, randomly changing this phase each time the unit is repeated. 
Our method is universal, that is, it can be directly incorporated to arbitrary DD sequences and is applicable for any physical realisation of a qubit sensor.

\begin{figure}
\center
\includegraphics[width=0.95\columnwidth]{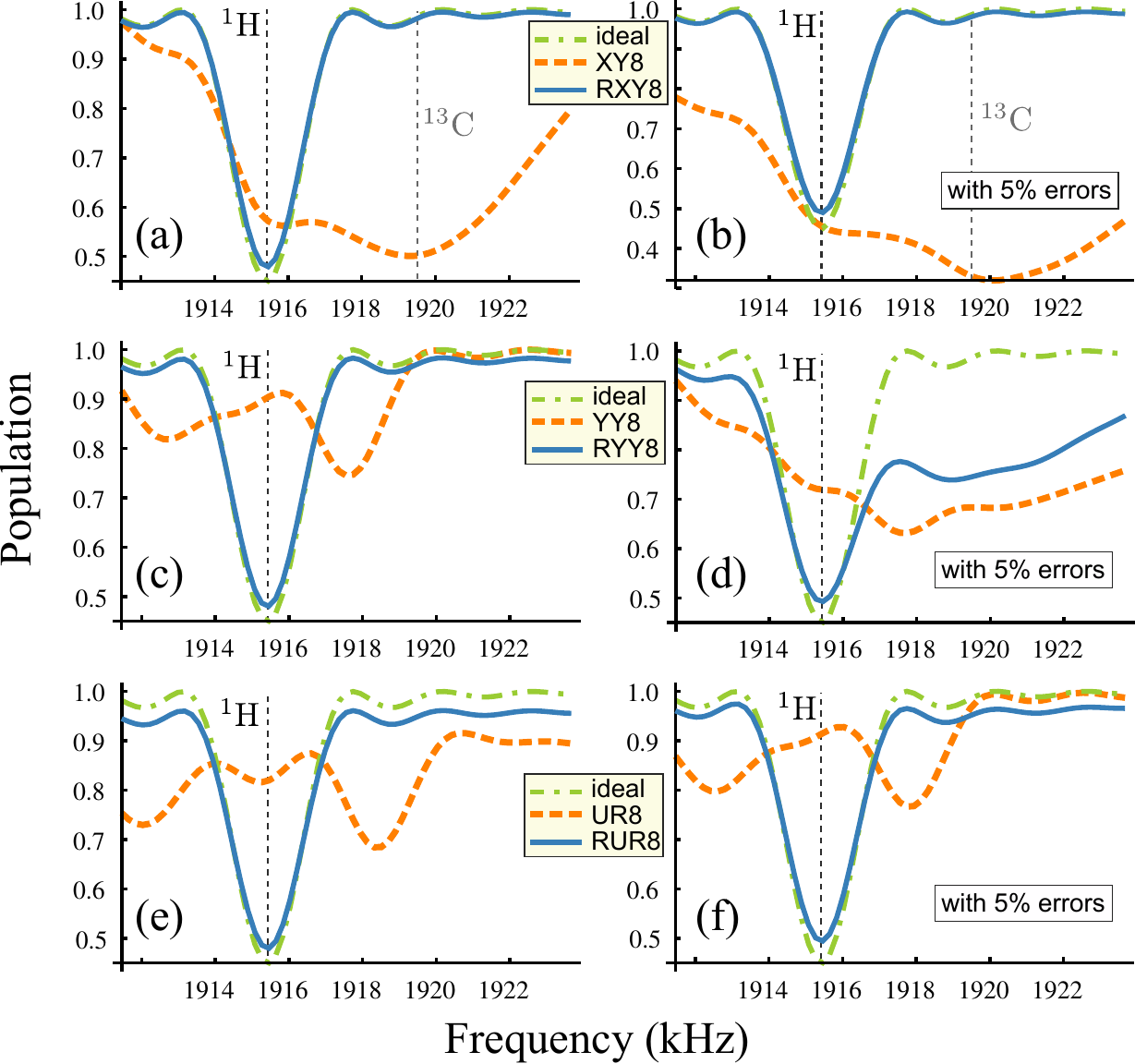}
\caption{Quantum spectroscopy with DD.
(a) Simulated averaged population signal
as a function of the DD frequency [$1/(2\tau)$ for pulse spacing $\tau$]. One $^{1}{\rm H}$ spin and one $^{13}{\rm C}$ spin are coupled to the NV centre via the hyperfine-field components~\cite{casanova2015robust,seeSM} $(A_{\perp},A_{\parallel})=2\pi\times(2,1)$ kHz and  $2\pi\times(5,50)$ kHz, respectively. The orange dashed line (blue solid line) is the signal obtained by a standard XY8 (randomised XY8) sequence using rectangular $\pi$ pulses with a time duration of $200$ ns and $M=200$. The presence of the $^{13}{\rm C}$ distorts the proton spin signal centred at the proton spin frequency (see the vertical dashed lines indicating the target $^{1}{\rm H}$ and the spurious $^{13}{\rm C}$ resonance frequencies for a magnetic field 450~G). The randomised XY8 sequence significantly reduces the signal distortion due to non-instantaneous control and reveals the real proton signal (see the green dash-dotted line for the signal obtained by a perfect XY8 sequence).
(b) As (a) but adding $5\%$ (in terms of the ideal Rabi frequency) of errors in both driving amplitude and frequency detuning to the $\pi$ pulses. (c) and (d) [(e) and (f)] are the same as (a) and (b) but for the YY8~\cite{shu2017unambiguous} [UR8~\cite{genov2017arbitrarily}] sequence. Despite the YY8 sequence - which uses single-axis control to mitigate the spurious peak in the $^{13}{\rm C}$ spectrum when there is no pulse error - the presence of the $^{13}{\rm C}$ still distorts the proton spin signal centred at the proton spin frequency. 
In all cases, the randomised protocol reduces the signal distortion due to non-instantaneous control and control errors.}
\label{FigH}
\end{figure}

\emph{DD-based quantum sensing.--}
Whilst our method is applicable to any qubit sensor, we illustrate it here with single NV centres. For all experiments in this work a bias magnetic field between 400 and 500 Gauss aligned with the NV-axis splits the degenerate $m_s=\pm 1$ spin states allowing the selective addressing of the $m_s=0 \leftrightarrow m_s=-1$ transition, which represents our sensor qubit with the qubit states $|0(1)\rangle$ [see Fig.~\ref{Fig1}(a) and \cite{seeSM} for details of the experimental set-up]. 
The sensor qubit and its environmental interaction takes the general form $\hat{H}^\prime(t)=\frac{1}{2}\hat{\sigma}_{z}\hat{E}(t)$. Here $\hat{\sigma}_{z}=|0\rangle\langle 0|-|1\rangle\langle 1|$ is the Pauli operator of the sensor qubit, and $\hat{E}(t)$ is an operator that includes the target signal which oscillates at a particular frequency as well as the presence of noisy environmental fluctuations. In the case of nuclear-spin sensing, $\hat{E}(t)$ contains target and bath nuclear spin operators oscillating at their Larmor frequencies. For AC magnetometry, $\hat{E}(t)$ describe classical oscillating magnetic fields. The aim of quantum sensing is to detect a target 
such as a single nuclear spin via the control of the quantum sensor with a sequence of DD $\pi$ pulses. The latter often corresponds to a periodic repetition of a basic pulse unit which has a time duration $T$ and a number $N$ of pulses [see Fig.~\ref{Fig1}(b)]. The propagator of a single $\pi$ pulse unit in a general form reads $\hat{U}_{\rm unit}(\{\phi_{j}\}) = \hat{\mathtt{f}}_{N+1}\hat{P}(\phi_N)\hat{\mathtt{f}}_{N}\cdots \hat{P}(\phi_2)\hat{\mathtt{f}}_{2}\hat{P}(\phi_1)\hat{\mathtt{f}}_{1}$,
where $\hat{\mathtt{f}}_{j}$ are the free evolutions separated by the control $\pi$ pulses with the 
propagator $\hat{P}(\phi_j)$. Errors in the control are included in  $\hat{P}(\phi_j)$, while the 
different pulse phases $\phi_j$ are used by robust DD sequences to mitigate 
the effect of detuning and amplitude errors of the $\pi$ pulses. Using $M$ repetitions of the 
basic DD unit [see Fig.~\ref{Fig1}(c) for the case of a standard construction] allows for $M$-fold
increased signal accumulation time $T_{\rm total}=M T$ which enhances the acquired contrast of the weak
signal as $\propto M^2$~\cite{zhao2011atomic} and improves the fundamental frequency resolution to $\sim 1/T_{\rm total}$.

To see how a target signal is sensed, we write the Hamiltonian $H^\prime(t)$ in the interaction picture of the DD control as~\cite{seeSM}
\begin{equation}
 \hat{H}(t) = \frac{1}{2}F_z(t) \hat{\sigma}_z \hat{E}(t) + \frac{1}{2}[F_{\perp}(t) \hat{\sigma}_{-}+{\rm H.c.}]\hat{E}(t), \label{HInt}
\end{equation}
where $\hat{\sigma}_{-}=|0\rangle\langle 1|$. For ideal instantaneous $\pi$ pulses, $F_{\perp}(t)=0$ vanishes [see Fig.~\ref{Fig1} (b) which shows how the $F_{\perp}(t)$ vanishes between the $\pi$ pulses] and the modulation function $F_{z}(t)$ is the stepped modulation function widely used in the literature, that is, $F_{z}(t)=(-1)^{m}$ when $m$ $\pi$-pulses have been applied up to the moment $t$. The role of a DD based quantum sensing sequence is to tailor $F_{z}(t)$ such that it oscillates at the same frequency as the target signal in $\hat{E}(t)$, allowing resonant coherent coupling between the sensor and the target.

\begin{figure}
\center
\includegraphics[width=0.9\columnwidth]{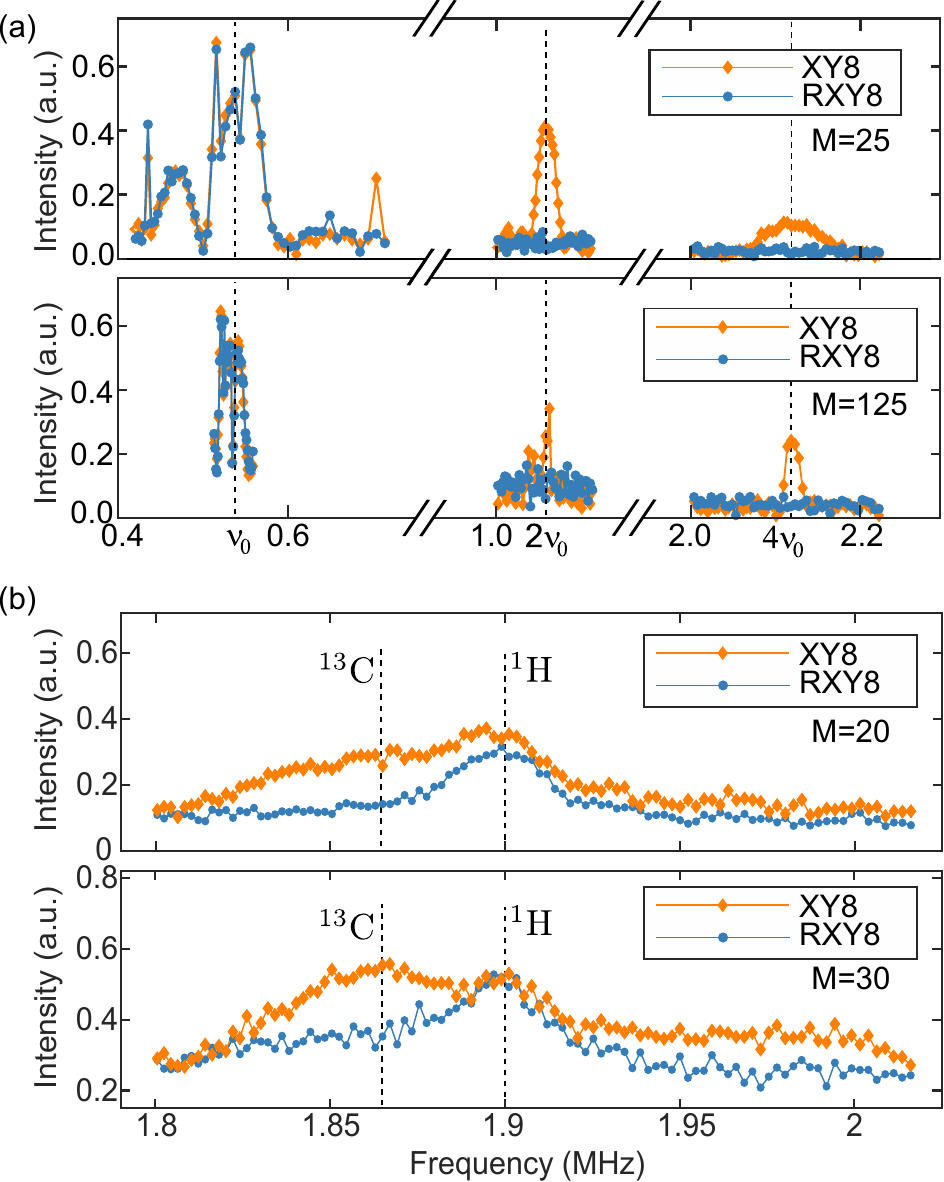}
\caption{Removing spurious response with the phase randomisation protocol. (a) In the measured spectrum of an AC magnetic field sensed by a standard repetition of the XY8 sequence (see orange diamonds), the non-instantaneous $\pi$ pulses produce spurious peaks at the frequencies $2\nu_{0}$ and $4\nu_{0}$. Repeating the XY8 sequence with phase randomisation (see blue bullets) preserves the desired signal centred at $\nu_{0}$ and efficiently suppresses all the spurious peaks. The XY8 unit was repeated $M=25$ times in the upper panel and $M=125$ in the lower panel for a longer sensing time. (b) Detection of proton spins using the XY8 sequence. For the measured spectrum obtained by the standard protocol, the $^{13}{\rm C}$ nuclear spins naturally in diamond produce a strong and wide spurious peak that hinders proton spin detection. Using the randomisation protocol, the spurious $^{13}{\rm C}$ peak has been suppressed, revealing the proton spin signal centred around a frequency of 1.9 MHz.}
\label{FigS}
\end{figure}

In realistic situations, where the $\pi$ pulses are not instantaneous due to limited control power, 
the function $F_{\perp}(t)$ has a non-zero value during $\pi$ pulse execution and $F_z(t)$ 
deviates from $\pm 1$~\cite{lang2017enhanced,lang2019non} [see Fig.~\ref{Fig1} (b) for the example 
of XY8 sequences]. While it is possible to eliminate the effect of deviation in $F_z(t)$ by pulse 
shaping technique~\cite{casanova2018shaped}, the presence of non-zero $F_{\perp}(t)$ may still 
alter the expected signal or cause spurious peaks to appear~\cite{loretz2015spurious}. In general, 
an oscillating component with a frequency $k/T_{\rm total}$ ($k$ being an integer) in $\hat{E}(t)$, not resonant with $F_z(t)$,
will create spurious response when the Fourier amplitude~\cite{lang2017enhanced,seeSM}
\begin{equation}
f^{\perp}_{k}=\frac{1}{T_{\rm total}}\int_{0}^{T_{\rm total}}F_{\perp}(t)\exp(-i 2\pi k t/T_{\rm total})dt
\end{equation}
of $F_{\perp}(t)$ is non-zero. 
This spurious response can cause false signal identification, e.g., a wrong conclusion on the  detected nuclear species~\cite{loretz2015spurious}, exemplified in Figs.~\ref{FigH} and \ref{FigS}. 
Suppressing the spurious response from $^{13}\rm{C}$ nuclei is especially critical, as it allows reliable nanoscale NMR or MRI without the use of hard to manufacture and consequently expensive, highly isotopically $^{12}\rm{C}$ purified diamond. However, as shown in Fig.~\ref{FigH} (c),(d), even
for a YY8 sequence (designed to remove spurious resonances~\cite{shu2017unambiguous}) the target proton signal is still perturbed by other nuclear species ($^{13}\rm{C}$ in this case).
In the presence of amplitude and detuning errors, standard strategies perform even worse.

To remove all spurious peaks, one seeks to design a DD sequence that minimises the effect of $F_{\perp}(t)$ in a robust manner. We observe that by introducing a global phase to all the $\pi$ pulses, the form of $F_{z}(t)$ is unchanged but a phase factor is added to $F_{\perp}(t)$. This motivates the following method to preserve $F_{z}(t)$ and to suppress the effect of $F_{\perp}(t)$ by phase randomisation.

\emph{Phase randomisation.--}
In the randomisation protocol, a random global phase $\Phi_{r,m}$ (where the subscript $r$ means a random value) is added to all the pulses within each unit $m$, as shown in Fig.~\ref{Fig1}(d). The propagator of $M$ DD units with independent global phases reads
$\hat{U}_{r} = \prod_{m=1}^{M}\hat{U}_{\rm unit}(\{\phi_j+\Phi_{r,m}\})$.
If one sets all the random phases $\Phi_{r,m}$ to the same value (e.g. zero) the original DD sequence can be recovered [Fig.~\ref{Fig1}(c)]. Since each of the global phases does not change the internal structure (i.e., the relative phases among $\pi$ pulses) of the basic unit, the robustness of the basic DD sequence is preserved. On the other hand, as we will show in the following, these random global phases prevent control imperfections from accumulating. 

\emph{Universal suppression of spurious response.--} The randomisation protocol provides a universal method to suppress spurious response. For the sequence with randomisation, one can find that the Fourier amplitude reads $f^{\perp}_{k}=Z_{r,M} \tilde{f}^{\perp}_{k/M}$, where $\tilde{f}^{\perp}_{k/M}=\frac{1}{T}\int_{0}^{T}F_{\perp}(t)\exp(-i\frac{2\pi k t}{M T})dt$ is the Fourier component defined over a single period $T$~\cite{seeSM}. For random phases $\{\Phi_{r,m}\}$, the factor
\begin{equation}
Z_{r,M} =\frac{1}{M} \sum_{m=1}^{M} \exp(i \Phi_{r,m}), \label{eq:phaseAverage}
\end{equation}
captures the effect of the randomisation protocol. Due to the random values of the phases $\Phi_{r,m}$, $Z_{r,M}$ becomes a (normalised) 2D random walk with $\langle|Z_{r,M}|^2\rangle=1/M$ thus suppressing the contrast of spurious response by a factor of $1/(2M)$ compared with the standard protocol~\cite{seeSM}. 
Here, we note that one can design a set of specific (i.e. not random) phases $\Phi_{r,m}$ that minimise a certain $f^{\perp}_{k}$ completely. However, this set of phases would be specific to one $k$-value (i.e. it does not suppress all spurious peaks simultaneously). In this respect, the power of our method is that it is simple to implement and fully universal, suppressing all spurious peaks produced by any sequence whilst still retaining the ideal signal, as shown in Fig.~\ref{FigH}.

To experimentally benchmark the performance, we carried out nanoscale detection of a classical AC magnetic field  [Fig.~\ref{FigS} (a)] and, separately, the nanoscale NMR detection of an ensemble of proton spins with a natural $^{13}$C abundance ($1.1\%$) diamond [Fig.~\ref{FigS} (b)]. The standard repetition of the XY8 sequence, which was widely used in various sensing and sensing based applications (e.g., see Refs.~\cite{staudacher2013nuclear,rugar2015proton,deVience2015nanoscale,shi2015single,glenn2018high,abobeih2018one,humphreys2018deterministic,rosskopf2017quantum,loretz2015spurious,pham2016nmr}),
produces spurious peaks when the duration of $\pi$ pulses is non-zero. In contrast, the randomisation protocol suppresses all the spurious peaks in the spectrum efficiently, and the spurious background noise from a $^{13}{\rm C}$ nuclear spin bath in diamond was removed while the desired proton signal was unaffected, demonstrating a clear and unambiguous proton spin detection without the use of $^{12}$C isotopically pure diamonds.

In the experiments, we have repeated the randomisation protocol with $K=10$ samples of the random phase sequences $\{\Phi_{r,m}\}$ and averaged out the measured signals. 
This reduces the fluctuations of the (suppressed) spurious peaks, introduced by the applied random phases, because the variance of $|Z_{r,M}|^2$ (which is $(M-1)/M^3$) is further reduced by a factor of $1/K$~\cite{seeSM}.

Removing the spurious response also improves the accuracy, for example, in measuring the depth of individual NV centres~\cite{pham2016nmr}. By falsely assuming that all the signal around $1.9$ MHz  obtained by the standard XY8 sequences originates from hydrogen spins, the computed NV centre depth would be $5.88\pm0.52$~nm, instead of $7.62\pm0.29$~nm obtained by the randomised XY8 - a deviation of about 30~$\%$ [see Fig.~\ref{FigS} (b)]. 

\begin{figure}
\center
\includegraphics[width=1.0\columnwidth]{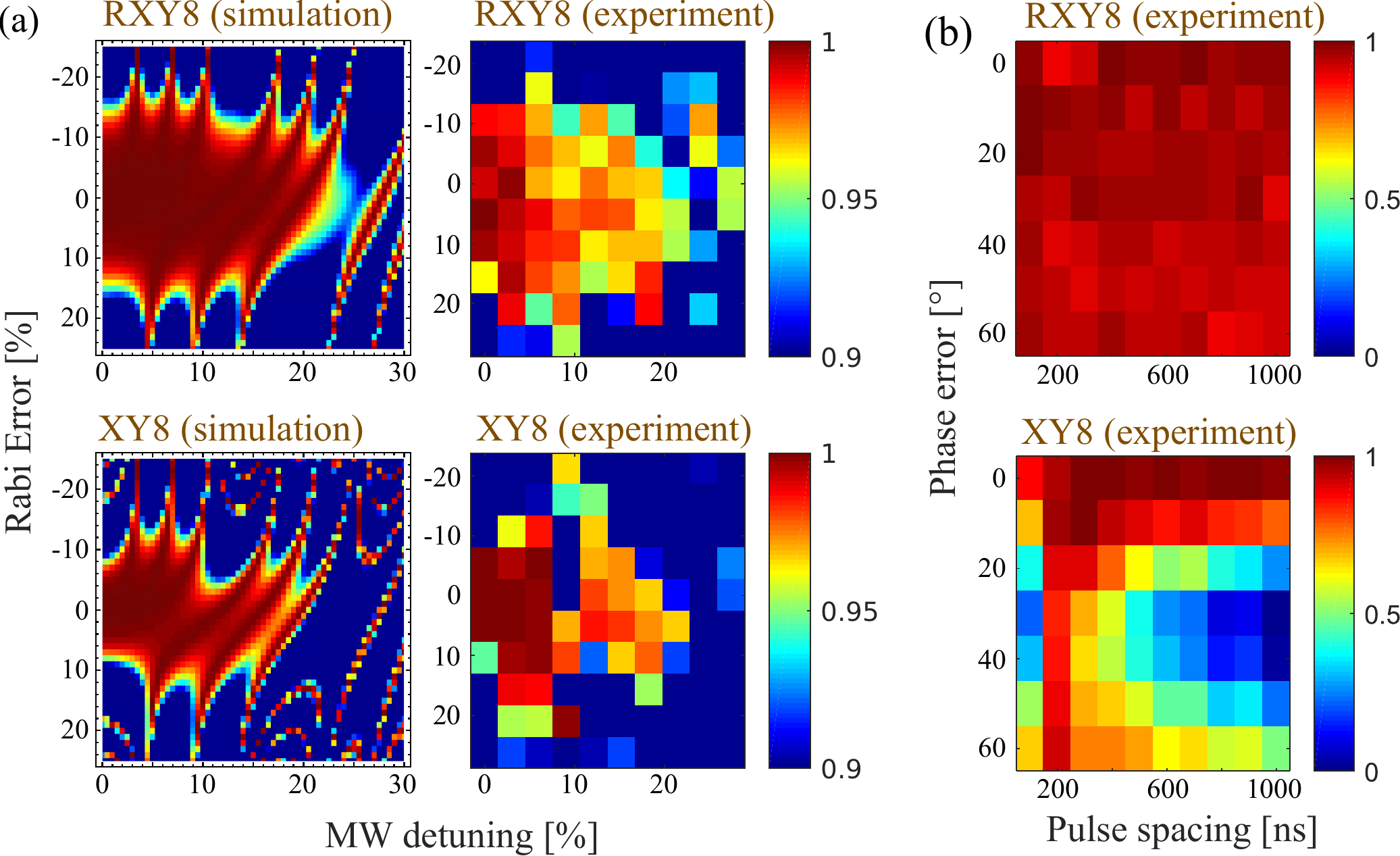}
\caption{Experimental enhancement of sequence robustness with the phase randomisation protocol. (a) The fidelity of XY8 sequences as a function of detuning and Rabi frequency errors for randomisation (upper panels) and standard (lower panels) protocols. The control errors are measured in terms of the ideal Rabi frequency $\Omega_{\rm ideal}= 2\pi\times 32.8 $ MHz. The sequences have inter-pulse spacing $200$ ns and $M=25$ XY8 units. (b) The fidelity of XY8 sequences with respect to a static phase error between the X and Y pulses and the inter-pulse time interval $\tau$, for randomised (upper panels) and standard (lower panels) protocols with $M=12$. Resonant microwave $\pi$ pulses are used with a Rabi frequency $\Omega_{\rm ideal}= 2\pi\times 66.6 $ MHz. 
}
\label{FigR}
\end{figure}

\emph{Enhancement on control robustness.--} 
As indicated in Fig.~\ref{FigH}, the randomisation protocol also enhances the robustness of the whole DD sequence. For simplicity, in the following discussion we neglect the effect of the environment and concentrate on static control imperfections. 
The latter introduce errors in the form of non-zero matrix elements 
$\langle 0|\hat{U}_{\rm unit}|1\rangle = C \epsilon + O(\epsilon^2)$, where $\epsilon$ is a small parameter and $C$ is a prefactor depending on the explicit form of control (see~\cite{seeSM} for details).
For the standard protocol where the same $\hat{U}_{\rm unit}$ block is repeated, the static errors accumulate coherently, yielding $\langle 0|(\hat{U}_{\rm unit})^{M}|1\rangle = MC \epsilon+ O(\epsilon^2)$. The random phases in the randomisation protocol avoids this coherent error accumulation and one can find $\langle 0|\hat{U}_{r}|1\rangle = Z_{r,M} MC \epsilon+ O(\epsilon^2)$, where the error is suppressed by the factor $Z_{r,M}$ which is given by Eq.~(\ref{eq:phaseAverage}) for random phases~\cite{seeSM}. Compared with the suppression of control imperfections by deterministic phases, the randomisation protocol is universal and achieves both suppression of spurious response and enhancement of robustness, without loss of sensitivity to target signals as shown in Figs.~\ref{FigH} and \ref{FigS}.

In Fig.~\ref{FigR} (a), we show the robustness of the widely used XY8 sequence, with respect to amplitude bias and frequency detuning of the microwave pulses, for the randomisation and standard protocols. The simulation and experiment demonstrate robustness improvement after applying phase randomisation.  As shown in Fig.~\ref{FigR} (b), the randomisation protocol also suppresses errors in pulse phases. The latter is especially relevant for digital pulsing devices where the signal from a microwave source is split-up and the phase in one arm is shifted by suitable equipments. On top of errors due to the working accuracy of these devices, different cable lengths in both arms can sum up to errors in the relative phase.

\emph{Conclusion.--}
We present a randomisation protocol for DD sequences that efficiently and universally suppresses spurious response whilst maintaining the expected signal. This method is simple to implement, only requiring additional random control-pulse phases, and is valid for all DD sequence choices. The protocol functions equally well for quantum and classical signals, allowing clear and unambiguous AC field and nuclear spin detection, e.g., with the widely used XY family of sequences. Furthermore, the protocol also enhances the robustness of the whole pulse sequences. 
For sensing experiments with NV centres, the protocol reduces the reliance on hard to manufacture, expensive, highly isotopically purified diamond.
The method has a general character being equally applicable to other quantum platforms and other DD applications. For example, it could be used to improve correlation spectroscopy~\cite{laraoui2013,ma2016proposal,wang2017delayed,rosskopf2017quantum} in quantum sensing and fast quantum gates in trapped ions~\cite{arrazola2018arrazola,manovitz2017fast} where DD has been used as an important ingredient.

\emph{Acknowledgements.--}
M.~B.~P. and Z.-Y.~W. acknowledge support by the ERC Synergy grant BioQ (Grant No. 319130), the EU project HYPERDIAMOND and AsteriQs, the QuantERA project NanoSpin, the BMBF project DiaPol, the state of Baden-W{\"u}rttemberg through bwHPC, and the German Research Foundation (DFG) through Grant No. INST 40/467-1 FUGG. J.~E.~L. is funded by an EPSRC Doctoral Prize Fellowship. F.~J., S.~S., L.~M., and J.~L. acknowledge support of Q-Magine of the QUANTERA, DFG (FOR 1493, SPP 1923, JE 290/18-1 and SFB 1279), BMBF (13N14438, 16KIS0832, and 13N14810), ERC (BioQ 319130), VW Stiftung and Landesstiftung BW. J.~C. acknowledges financial support from Juan de la Cierva Grant No. IJCI-2016-29681.

\pagebreak{}\clearpage{}

\begin{center}
\textbf{\large{{Supplemental Material}}}{\large{ }}
\par\end{center}{\large \par}

\setcounter{equation}{0} \setcounter{figure}{0} \setcounter{table}{0}
\setcounter{page}{1} \makeatletter \global\long\def\theequation{S\arabic{equation}}
 \global\long\def\thefigure{S\arabic{figure}}
 \global\long\def\bibnumfmt#1{[S#1]}
 \global\long\def\citenumfont#1{S#1}

\section{Experimental methods}

\subsection{Diamonds}

All experiments were performed on single NV centres.
For the nanoscale NMR experiments [Fig. 3(b) of the main text] a $^{13}$C natural abundance diamond was implanted with $^{15}N$ ions using an energy of 1.5\,keV and a dose of $2 \times 10^9\mathrm{\frac{^{15}N^+}{cm^2}}$. 
Subsequent annealing in vacuum at 1000$^\circ$C for 3 hours created shallow single NV centres with depths around $5\pm 1$ \,nm.
For the experiments measuring the classical AC fields [Fig. 3 (a)] we used a different diamond, which was polished into a solid immersion lens.
In order to create NV centres in this diamond, the flat surface was overgrown with an about 100\,nm thick layer of isotopically enriched $^{12}$C (99.999$\%$) using the plasma enhanced chemical vapor deposition method, with parameters as in \cite{sm:Osterkamp}. 
The same diamond was used for the experiments showing the improved robustness of the randomisation protocol (Fig. 4).
The experiments presented in \ref{add_exp} were measured with an about 4\,$\mu$m deep NV in a flat diamond with 0.1\% $^{13}$C content.
Before experiments, all diamonds were boiled in a 1:1:1 tri-acid mixture (H$_2$SO$_4$:HNO$_3$:HClO$_4$) for 4 hours at 130$^{\circ}$C. 

\subsection{Setup}

Using a home-built confocal setup, read-out and initialisation (into the $|0\rangle$ spin state) of the NV center was performed using a 532\,nm laser. 
The laser beam was chopped using an acousto optical modulator into pulses of 3\,$\mu$s duration.
The spin-dependent fluorescence from the NV spin states was detected using an avalanche photodiode. The first 500\,ns of the every laser pulse yield the spin population while the fluorescence between 1.5\,$\mu$s and 2.5\,$\mu$s was used to normalise the data. 
Magnetic bias fields between 400\,G and 500\,G were used to lift the degeneracy of the $|-1\rangle, |+1\rangle$ spin states and create an effective qubit.

Microwave pulses resonant with the NV centre spin were applied using a 20\,$\mu$m diameter copper wire placed on the diamond surface as an antenna.
The pulses were generated with an Arbitrary Waveform Generator (Tektronix AWG70001A, sampling rate 50GSamples/s) and amplified 
to give Rabi frequencies between 5-70\,MHz.
The same wire was used to apply classical radio-frequency fields generated by a Gigatronics 2520B signal generators. 
For the classical AC field detection, background magnetic noise at the frequencies detected was determined to be at least 100 fold weaker than the measured signals. 

\subsection{Measurement protocol}

All experiments were performed using the QuDi software suite \cite{sm:qudi}. 
For the randomised protocols the standard versions were modified by adding a random global phase to all $\pi$ pulses in a basic unit, as described in the main text. These phases were generated using the Python package 'random' with a uniform distribution between 0 and 2$\pi$. 
Before applying the dynamical decoupling protocols, the spin of the NV centre is initialized in a coherent superposition ($\frac{1}{\sqrt{2}}(|0\rangle+|1\rangle$)). Therefore, additionally to the laser pulse, a $\pi$/2 pulse is applied.
Also, before the readout the acquired sensor phase is mapped into a population difference by an additional $\pi$/2 pulse.
For our experiments showing the improved robustness we intentionally introduce pulse errors to the $\pi$ pulses. 
Those errors were calibrated in terms of the real Rabi frequency. 
Thereby, the two $\pi$/2 pulses were always applied error-free.
Every experiment was repeated several times under identical experimental conditions, but with different sets of phases, and the resulting data was averaged.

\section{Hamiltonian under dynamical decoupling control }

As stated in the main text, the Hamiltonian without dynamical decoupling (DD) control has the general form 
\begin{equation}
\hat{H}^{\prime}(t)=\frac{1}{2}\hat{\sigma}_{z}\hat{E}(t),
\end{equation}
where $\hat{\sigma}_{z}=|0\rangle\langle0|-|1\rangle\langle1|$ is the Pauli operator of the sensor. The environment operator $\hat{E}(t)$ includes both the target signal to be sensed and environmental noise. For the relevant case of nuclear spin sensing, $\hat{E}(t)=\frac{1}{2}\sum_{n}\left[\left(A_{n}^{\perp}\hat{I}_{n}^{+}e^{-i\omega_{n}t}+{\rm h.c.}\right)+A_{n}^{\parallel}\hat{I}_{n}^{z}\right],$where
$\hat{I}_{n}^{\alpha}$ ($\alpha=x,y,z$) are spin operators for the $n$th nuclear spin. $A_{n}^{\perp}$ and $A_{n}^{\parallel}$ are components of hyperfine field at the position of the nuclear spin. The nuclear spin precession frequency $\omega_{n}$ is the Larmor frequency of the nuclear spin shifted by the hyperfine field at the location of the nuclear spin. For the case of a classical AC field, $\hat{E}(t)$ takes the form $\sum_{n}b_{n}\cos(\omega_{n}t+\phi_{n})$. 

A sequence of applied microwave pulses yields the control Hamiltonian
\begin{equation}
\hat{H}_{{\rm ctrl}}(t)=\frac{1}{2}\Omega(t)\left[\hat{\sigma}_{x}\cos\phi(t)+\hat{\sigma}_{y}\cos\phi(t)\right].
\end{equation}
In the rotating frame with respect to the control $\hat{H}_{{\rm ctrl}}(t)$, the Hamiltonian $\hat{H}^{\prime}(t)$ becomes 
\begin{equation}
\hat{H}(t)=\frac{1}{2}\hat{\sigma}(t)\hat{E}(t),
\end{equation}
where $\hat{\sigma}(t)$ is $\hat{\sigma}_{z}$ in the Heisenberg picture with respect to $\hat{H}_{{\rm ctrl}}(t)$. In the following, we derive $\hat{\sigma}(t)$ and hence Eq. (2) in the main text. 

If a $\pi$ pulse is applied at time $t_j$, the evolution driven by $\hat{H}_{{\rm ctrl}}(t)$ reads $\hat{P}_{j}(\theta) = \exp\left[-i\frac{1}{2} \theta \left( \hat{\sigma}_{x} \cos\phi_{j} + \hat{\sigma}_{y} \cos\phi_{j} \right) \right]$,
where $\theta=\theta(t)\in[0,\pi]$ is the angle of rotation and $\phi(t_j)=\phi_j$. Defining $\hat{P}_{j}(\pi)\equiv \hat{P}_{j}$ as the propagator of a single $\pi$ pulse, the propagator for $2n+1$ ($j=0,1,\ldots$) pulses
\begin{align}
\hat{U}_{2n+1} & =\hat{P}_{2n+1}\cdots\hat{P}_{2}\hat{P}_{1}\\
 & =(-1)^{n+1}e^{i\frac{\pi}{2}}\exp\left(-i\varphi_{2n+1}\right)|0\rangle\langle1|+{\rm h.c}.,
\end{align}
and for $2n$ pulses
\begin{align}
\hat{U}_{2n} & =\hat{P}_{2j}\cdots\hat{P}_{2}\hat{P}_{1}\\
 & =(-1)^{n}\exp\left(i\varphi_{2n}\right)|0\rangle\langle0|+(-1)^{n}\exp\left(-i\varphi_{2n}\right)|1\rangle\langle1|,
\end{align}
where $\varphi_{2n+1}=-\sum_{l=1}^{2n+1}(-1)^{l}\phi_{l}$ and $\varphi_{2n}=-\sum_{l=1}^{2j}(-1)^{l}\phi_{l}$.
Using $\hat{U}_{2n+1}$ and $\hat{U}_{2n}$, we find $\hat{\sigma}_{z}$
in the rotating frame of the control during the $j$th pulse
\begin{align}
\hat{\sigma}(t) & =[\hat{P}_{j}(\theta)\hat{U}_{j-1}]^{\dagger}\hat{\sigma}_{z}[\hat{P}_{j}(\theta)\hat{U}_{j-1}]\\
 & =F_{z}(t)\hat{\sigma}_{z}+\left[F_{\perp}(t)|1\rangle\langle0|+{\rm h.c.}\right],
\end{align}
where the modulation functions are
\begin{equation}
F_{z}(t)=(-1)^{j-1}\cos\theta\label{sm:eq:Fz}
\end{equation}
\begin{equation}
F_{\perp}(t)=i(-1)^{j-1}\exp\left\{-i[2\sum_{l=1}^{j-1}(-1)^{l}\phi_{l}+(-1)^{j}\phi_{j}]\right\}\sin\theta.\label{sm:eq:Fx}
\end{equation}
Because $\theta=\theta(t)$ in Eqs. (\ref{sm:eq:Fz}) and (\ref{sm:eq:Fx}) is the pulse area that the $j$th pulse has rotated at the moment $t$, for instantaneous pulses $F_{\perp}(t)$ has no effect (because $\sin\theta=0$ for all time $t$) and $F_{\perp}(t) \in \{\pm 1\}$. For the realistic case that the pulses are not instantaneous, $F_{\perp}(t)$ is non-zero during the $\pi$ pulses. 

\section{Fourier amplitudes of the modulation functions}
For DD sequences that are $M$ periodic repetitions of a basic pulse unit with period $T$ and $F_{\alpha}(t+T)=F_{\alpha}(t)$ ($\alpha=z,\perp$),
the $k$th Fourier amplitude of the modulation functions (over the total sequence time $T_{{\rm total}}=MT$) is 
\begin{align}
f^{\alpha}_{k} & \equiv\frac{1}{MT}\int_{0}^{MT}F_{\alpha}(t)\exp\left(-i\frac{2\pi kt}{MT}\right)dt\\
 & =\frac{1}{MT}\sum_{m=1}^{M}\int_{(m-1)T}^{mT}F_{\alpha}(t)\exp\left(-i\frac{2\pi kt}{MT}\right)dt\\
 & =c_{k,M}\tilde{f}^{\alpha}_{k/M}
\end{align}
where
\begin{equation}
\tilde{f}^{\alpha}_{k/M} \equiv \frac{1}{T}\int_{0}^{T}F_{\alpha}(t)\exp\left(-i\frac{2\pi k t}{M T}\right) dt,
\end{equation}
and $c_{k,M}=\frac{1}{M}\sum_{m=1}^{M}\exp\left(-i\frac{2\pi k(m-1)}{M}\right)$. When $k/M$ is not an integer $Mc_{k,M}$ is a sum over roots of unity so it cancels to zero. When $k/M$ is an integer however the sum gives $c_{k,M}=1$. Therefore for standard repetitions of a basic pulse unit, we obtain (for $k=1,2,\ldots$)
\begin{equation}
f^{\alpha}_{k}=\begin{cases}
\tilde{f}^{\alpha}_{k/M} & {\rm if}\ k/M\in\mathbb{Z},\\
0 & {\rm otherwise}.
\end{cases}
\end{equation}

Under the randomisation protocol a random phase is added to all pulses in the $m$-th repetition of the basic unit, so a set of $M$ random phases is generated, $\{\Phi_{r,m}|m=1,\ldots,M\}$. This transformation does not affect $F_{z}(t)$ but alters $F_{\perp}(t)\rightarrow F_{\perp}(t)e^{i\Phi_{r,m}}$ for the $m$-th unit of the sequence. The Fourier amplitudes $f^{z}_{k}$ are thus unaffected but we have  
\begin{align}
f^{\perp}_{k} & =\frac{1}{MT}\int_{0}^{MT}F_{\perp}(t)\exp\left(-i\frac{2\pi kt}{MT}\right)dt\\
 & =\frac{1}{MT}\sum_{m=1}^{M}\int_{(m-1)T}^{mT}F_{\perp}(t)e^{-i\Phi_{r,m}}\exp\left(-i\frac{2\pi kt}{MT}\right)dt\\
 & =Z_{r,M}\tilde{f}^{\perp}_{k/M},
\end{align}
where $Z_{r,M}=\frac{1}{M}\sum_{m=1}^{M}e^{i[\Phi_{r,m}-2\pi k(m-1)/M]}$.
Because $\Phi_{r,m}$ is chosen randomly, $\Phi_{r,m}-2\pi k(n-1)/M$ is also random and we can write 
\begin{equation}
Z_{r,M}=\frac{1}{M}\sum_{m=1}^{M}\exp(i\Phi_{r,m}),
\end{equation}
which is Eq. (5) in the main text. Here $Z_{r,M}$ is a sum of random complex phases and represents a 2D random walk. It can be shown that $|Z_{r,M}|^2$ has the average $\langle|Z_{r,M}|^2\rangle=1/M$ and the variance $\langle(|Z_{r,M}|^2- \langle|Z_{r,M}|^2\rangle)^2\rangle =(M-1)/M^3$. For example, 
the average can be obtained as follows. By definition,
\begin{align}
|Z_{r,M}|^2 & = \frac{1}{M^2}\sum_{m,n=1}^{M}\exp[i(\Phi_{r,m}-\Phi_{r,n})] \\
& = \frac{1}{M^2} \left[M + \sum_{m\neq n}^{M}e^{i(\Phi_{r,m}-\Phi_{r,n})}\right].
\end{align}
Therefore, $\langle|Z_{r,M}|^2\rangle=1/M$ because the average of independent random phases is zero. Similarly, one obtains the variance of $|Z_{r,M}|^2$. 

Consider the signal of a single nuclear spin. The population signal of expected resonances is given by $P=\cos^2(\frac{1}{2} |f^{z}_{k}|A_{\perp}MT)$, where $A_\perp$ is the perpendicular coupling strength to a single spin-half~\cite{sm:lang2017enhanced}.  When the signals are weak this can be approximated by $P=1- (\frac{1}{2}|f^{z}_{k}|A_{\perp}MT)^2$ thus the signal contrast is proportional to $M^2$. This is unaffected by the addition of the random phase as $F_z(t)$ is insensitive to the pulse phases.

The spurious  signal of a nuclear spin is given by $P =1-\sin^2(\frac{1}{2}A_{\perp}|f^{\perp}_{k}|MT)\cos^2(\phi^{\perp}_{k})$, where $\phi^{\perp}_{k}$ is the complex phase of $f^{\perp}_{k}$~\cite{sm:lang2017enhanced}. For the standard protocol, we have $P= 1-\sin^2(\frac{1}{2}A_{\perp}|\tilde{f}^{\perp}_{k/M}|MT)\cos^2(\phi^{\perp}_{k})$. When the signal is weak this can be approximated by $P \approx 1-(\frac{1}{2}A_{\perp}|\tilde{f}^{\perp}_{k/M}|MT)^2\cos^2(\phi^{\perp}_{k})$ so when no random phase is added the spurious signal contrast is proportional to $M^2$. 
When the random phase is added the expected value of the signal contrast is given by $P \approx 1 - \frac{1}{8}M(T A_{\perp}|\tilde{f}^{\perp}_{k/M}|)^2$ (using $\langle |f^{\perp}_{k}|^2 \rangle = \langle |Z_{r,M}\tilde{f}^{\perp}_{k/M}|^2 \rangle = |\tilde{f}^{\perp}_{k/M}|^2/M$ and $\langle\cos^2(\phi_\perp^k)\rangle = 1/2$). Compared with standard repetitions of a basic pulse unit, this contrast only grows proportional to $M/2$ thus providing a significant suppression of spurious signals completely independent of the used pulse sequence. As shown above, the variance of the spurious signal due to random phases is determined by the variance of $|Z_{r,M}|^2$ (which is $(M-1)/M^3$). When one repeats the randomisation protocol with $K$ realizations of the random phase sequences $\{\Phi_{r,m}\}$ and average out the measure signals, the variance is further reduced by a factor of $1/K$ according to the central limit theorem.

\section{Enhancing sequence robustness}
For simplicity, in the following discussion we neglect the effect of the  environment and concentrate on static control imperfections. 

\subsection{Evolution operator of a basic pulse unit}

The evolution driven by a single $\pi$ pulse with control errors takes the general form
\begin{equation}
\hat{U}_{\pi}(\phi)=\left(\begin{array}{cc}
e^{-i\alpha}\sin\epsilon & ie^{-i(\beta+\phi)}\cos\epsilon\\
ie^{i(\beta+\phi)}\cos\epsilon & e^{i\alpha}\sin\epsilon
\end{array}\right).
\end{equation}
We assume that each pulse has the same static errors, that is, $\alpha$, $\beta$, $\epsilon$ are the same for all pulses. The pulse phase $\phi$ determined by the initial phase of the driving field is a controllable parameter. When $\epsilon = 0$ and $\beta=0$, $\hat{U}_{\pi}(\phi)$ describes a perfect $\pi$ pulse. 

Consider a basic unit with $N$ $\pi$ pulses applied at $t_{j}$ $(j=1,\ldots,N)$ with phases $\phi_{j}$. For simplicity, we use the transformation $t_{j+1}-t_{j}=\tau_{j}+\tau_{j+1}$ with $\tau_{0}\equiv0$. This transformation splits $t_{j+1}-t_{j}$ into two parts where  $\tau_{j}$ ($\tau_{j+1}$) is associate with the $j$th ($(j+1)$th) pulses. From the definition, we have
\begin{align}
\tau_{N+1} & =(t_{N+1}-t_{N})-\tau_{N}\\
 & =(-1)^{N}\sum_{j=0}^{N}(-1)^{j}(t_{j+1}-t_{j}).
\end{align}
by recursively using $\tau_{j+1}=(t_{j+1}-t_{j})-\tau_{j}$. Because a a basic DD unit is designed to eliminate static dephasing noise, the timing of the sequence satisfy $\sum_{j=0}^{N}(-1)^{j}(t_{j+1}-t_{j})=0$. In other words, $\tau_{N+1}=0$ for a basic pulse unit. 

With $\tau_{0}=\tau_{N+1}=0$ and that a detuning $\Delta$ of the control field introduces a control phase error $\Delta(t_{j+1}-t_{j})=\Delta(\tau_{j+1}+\tau_{j})$ during the times $t_{j}$ and $t_{j+1}$, the propagator of a basic pulse unit can be written as 
\begin{equation}
\hat{U}_{{\rm unit}}=\hat{U}_{N}\hat{U}_{N-1}\cdots \hat{U}_{2}\hat{U}_{1},
\end{equation}
by combining the contribution of a $\pi$ pulse and the free evolution we obtain
\begin{align}
\hat{U}_{j} & =\left(\begin{array}{cc}
e^{-i[\alpha+(\tau_{j}+\tau_{j-1})\Delta]}\sin\epsilon & ie^{-i[\beta+\phi_{j}-(\tau_{j}+\tau_{j-1})\Delta]}\cos\epsilon\\
ie^{i[\beta+\phi_{j}-(\tau_{j}+\tau_{j-1})\Delta]}\cos\epsilon & e^{i[\alpha+(\tau_{j}+\tau_{j-1})\Delta]}\sin\epsilon
\end{array}\right).\nonumber\\
 & =\left(\begin{array}{cc}
e^{-i[\alpha+(\tau_{j}+\tau_{j-1})\Delta]}\epsilon & ie^{-i[\beta+\phi_{j}-(\tau_{j}+\tau_{j-1})\Delta]}\\
ie^{i[\beta+\phi_{j}-(\tau_{j}+\tau_{j-1})\Delta]} & e^{i[\alpha+(\tau_{j}+\tau_{j-1})\Delta]}\epsilon
\end{array}\right)+O(\epsilon^2),
\end{align}

\begin{figure}[ht]
\begin{center}
\includegraphics[width=0.45\textwidth]{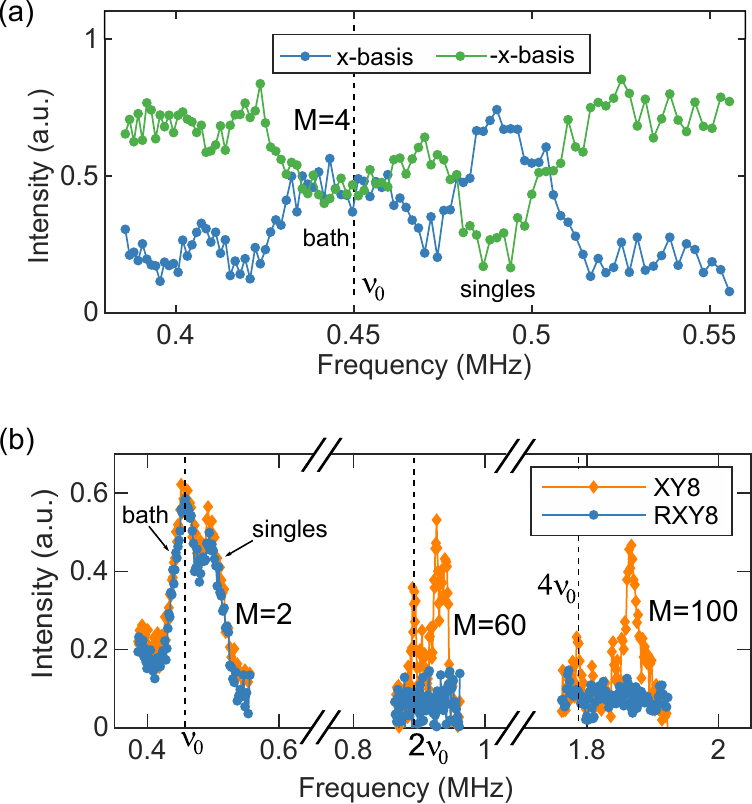}
\end{center}
\caption{Spectra of a single NV centre coupled to both individual $^{13}$C spins and the background $^{13}$C spin bath.
a) The readout in x and -x-bases highlights the saturation feature typical for a bath.
b) Comparison of standard XY8 and its randomisation version. The randomisation of the $\pi$ pulse phases suppresses the spurious signals efficiently.} 
\label{fig_bath_single}
\end{figure}

For two pulses, we find
\begin{equation}
\hat{U}_{j+1}\hat{U}_{j}=\left(\begin{array}{cc}
e^{i\varphi_{j}} & ic_{j}\epsilon\\
ic_{j}^{*}\epsilon & -e^{-i\varphi_{j}}
\end{array}\right)+O(\epsilon^{2}),
\end{equation}
where 
\begin{equation}
\varphi_{j}= \Delta(\tau_{j+1}-\tau_{j-1})-(\phi_{j+1}-\phi_{j})+\pi,
\end{equation}
and 
\begin{equation}
c_{j}=e^{-i[\beta+\phi_{j}+\alpha+\Delta(\tau_{j+1}-\tau_{j-1})]}+e^{-i[\beta+\phi_{j+1}-\alpha-\Delta(\tau_{j+1}+2\tau_{j}+\tau_{j-1})]},
\end{equation}
is a sum of phase factors where each term has a $\phi_{j}$ or $\phi_{j+1}$. Timing the $U_{j}$ recursively and using $\tau_{0}=\tau_{N+1}=0$, we obtain for even $N$
\begin{equation}
\hat{U}_{{\rm unit}}=\left(\begin{array}{cc}
e^{i\varphi} & iC\epsilon\\
iC^{*}\epsilon & e^{-i\varphi}
\end{array}\right)+O(\epsilon^{2}), \label{sm:eq:UunitEven}
\end{equation}
where 
\begin{equation}
\varphi=\sum_{j=1}^{N/2}\left[\phi_{2j-1}-\phi_{2j}+\pi\right],
\end{equation}
and $C$ is a sum of phase factors where each term has an independent sum of the phases $\phi_{j}$.
In deed, Eq.~(\ref{sm:eq:UunitEven})  has the general form of a pulse sequence with an even number of $\pi$ pulses with respect to the leading order error $\epsilon$~\cite{sm:genov2017arbitrarily}.

Similarly, we have for odd $N$, 
\begin{equation}
\hat{U}_{{\rm unit}}=\left(\begin{array}{cc}
C^{\prime*}\epsilon & ie^{-i(\varphi+\beta)}\\
ie^{i(\varphi+\beta)} & C^{\prime}\epsilon
\end{array}\right)+O(\epsilon^{2}),\label{sm:eq:UunitOdd}
\end{equation}
where 
\begin{equation}
\varphi=\sum_{j=1}^{(N-1)/2} \left[\phi_{2j-1}-\phi_{2j}+\pi\right]+\phi_{N},
\end{equation}
and $C^\prime$ is a sum of phase factors where each term has an independent sum of the phases $\phi_{j}$.

For the case that the lower-order errors of single $\pi$ pulses have been compensated by a robust sequence, one can still write the propagator in terms of the leading order error that has not been compensated by the sequence. The evolution operator of a single pulse sequence unit still has a general form given by Eq.~(\ref{sm:eq:UunitEven}) or (\ref{sm:eq:UunitOdd}), but may have another error $\epsilon_{\varphi}$ added to $\varphi$. For many sequences, such as the CP~\cite{sm:carr1954effects}, XY8~\cite{sm:gullion1990new}, AXY8~\cite{sm:casanova2015robust}, YY8~\cite{sm:shu2017unambiguous}, and UR-($4n+2$) ($n=1,2,\ldots$)~\cite{sm:genov2017arbitrarily} sequences, $\epsilon_{\varphi}$ is a higher-order error compared with $\epsilon$ and therefore can be neglected in the leading order error analysis. 

\subsection{Standard protocol}
It is obvious that the control errors coherently accumulate in the standard protocol where the basic pulse unit is repeated $M$ times as $\hat{U}  = (\hat{U}_{\rm unit})^M$. For example, for even $N$ and $\varphi=0$, 
\begin{equation}
\hat{U} = \left(\begin{array}{cc}
1 & i M C \epsilon \\
i M C^{*} \epsilon & 1
\end{array}\right)+O(\epsilon^{2}),
\end{equation}
where the error $MC\epsilon$ scales linearly with $M$. 

\subsection{Randomisation protocol}
When one adds a random global phase $\Phi_{r,m}$ on all the $\pi$ pulses in a basic DD unit, each $\hat{U}_{{\rm unit}}$ becomes
\begin{equation}
\hat{U}_{{\rm unit}}(\Phi_{r,m})=\left(\begin{array}{cc}
e^{i\varphi} & iCe^{-i\Phi_{r,m}}\epsilon\\
iC^{*}e^{i\Phi_{r,m}}\epsilon & e^{-i\varphi}
\end{array}\right)+O(\epsilon^{2}).
\end{equation}
For two $U_{{\rm unit}}$, we have 
\begin{equation}
\hat{U}_{{\rm unit}}(\Phi_{r,m+1})\hat{U}_{{\rm unit}}(\Phi_{r,m})=\left(\begin{array}{cc}
e^{2i\varphi} & iZ_m\epsilon\\
iZ_m^{*}\epsilon & e^{-2i\varphi}
\end{array}\right)+O(\epsilon^{2}),
\end{equation}
where $Z_m=e^{-i\varphi} C (e^{-i\Phi_{r,m+1}}+e^{-i(\Phi_{r,m}-2\varphi)})$ is a sum of two phase factors and can be equally written as $Z_j=e^{-i\varphi} C (e^{-i\Phi_{r,m+1}}+e^{-i\Phi_{r,m}})$ for random phases $\Phi_{r,m}$ and $\Phi_{r,m+1}$.
By mathematical induction, the evolution operator of $M$ DD units with random phases $\{\Phi_{r,m}\}$ is
\begin{align}
\hat{U}_{M} & =\hat{U}_{{\rm unit}}(\Phi_{r,M})\cdots \hat{U}_{{\rm unit}}(\Phi_{r,2})\hat{U}_{{\rm unit}}(\Phi_{r,1}),\\
 & =\left(\begin{array}{cc}
e^{iM\varphi} & i Z_{r,M}MC\epsilon\\
iZ_{r,M}^{*}MC^{*}\epsilon & e^{-iM\varphi}
\end{array}\right)+O(\epsilon^{2}),
\end{align}
where the error $MC\epsilon$ is suppressed by the factor $Z_{r,M}=\frac{1}{M}\sum_{m=1}^{M}\exp(i\Phi_{r,m})$ for the random phases $\{\Phi_{r,m}\}$. This result is valid for an odd number $N$ of pulses as well.

\section{Additional experiments}
\label{add_exp}
One of the most important advantages of quantum sensors is the possibility to measure quantum signals, such as hyperfine fields of single spins. 
This is highly relevant for the characterization of quantum systems. 
The randomisation protocol efficiently suppresses both spurious harmonics from a bath as well as from single spins.
In Fig. \ref{fig_bath_single}(a) we show the spectrum of an NV center that couples to both individual $^{13}$C spins and the background $^{13}$C spin bath.
The signal of the bath is centered around the bare Larmor of $^{13}$C at this bias field and shows the typical saturation highlighted by measuring the spectra for both x-basis and -x-basis readout.
The signal of at least one strongly coupled spin is shifted to higher frequencies due to the hyperfine coupling and it overlaps for the different readout bases.
In Fig. \ref{fig_bath_single}(b) we compare the spectra measured with standard XY8 and the randomisation version.
The identical signal shape and amplitude of the non-spurious signals verify that the randomized version does not alter the signal accumulation.
In order to amplify the spurious harmonics, we use larger number of $\pi$-pulses ($M$=60 and 100). 
We observe peaks at 2$\nu_0$ and 4$\nu_0$ for the standard XY8 method. 
For the same bias field the Larmor frequency of $^1$H is about 1.81\,MHz, what would make a differentiation very difficult. 
These spurious signals can be efficiently suppressed with the randomisation protocol.

\end{document}